\documentclass[%
reprint,
superscriptaddress,
nolongbibliography,
amsmath,amssymb,aps, 
prl,
]{revtex4-2}

\usepackage[utf8]{inputenc}
\usepackage[version=3]{mhchem}
\usepackage{graphicx}
\usepackage{dcolumn}
\usepackage{bm}
\usepackage{float}
\usepackage[usenames,dvipsnames]{color}

\begin{document}


\title{Direct visualization of surface spin-flip transition}

\author{Wenbo Ge}
\affiliation{Department of Physics and Astronomy, Rutgers University, Piscataway, New Jersey 08854, USA}
\author{Jinwoong Kim}
\affiliation{Department of Physics and Astronomy, Rutgers University, Piscataway, New Jersey 08854, USA}
\author{Ying-Ting Chan}
\affiliation{Department of Physics and Astronomy, Rutgers University, Piscataway, New Jersey 08854, USA}
\author{David Vanderbilt}
\affiliation{Department of Physics and Astronomy, Rutgers University, Piscataway, New Jersey 08854, USA}
\author{Jiaqiang Yan}
\affiliation{Materials Science and Technology Division, Oak Ridge National Laboratory, Oak Ridge, Tennessee 37831, USA}
\author{Weida Wu}
\email{wdwu@physics.rutgers.edu}
\affiliation{Department of Physics and Astronomy, Rutgers University, Piscataway, New Jersey 08854, USA}%

\begin{abstract}
We report direct visualization of spin-flip transition of the surface layer in antiferromagnet \ce{MnBi4Te7}, a natural superlattice of alternating \ce{MnBi2Te4} and \ce{Bi2Te3} layers, using cryogenic magnetic force microscopy (MFM). The observation of magnetic contrast across domain walls and step edges confirms that the antiferromagnetic order persists to the surface layers. The magnetic field dependence of the MFM images reveals that the surface magnetic layer undergoes a first-order spin-flip transition at a magnetic field that is lower than the bulk transition, in excellent agreement with a revised Mills’ model. Our analysis indicates an enhancement of the order parameter in the surface magnetic layer, implying robust ferromagnetism in the single-layer limit. The direct visualization of surface spin-flip transition not only opens up exploration of surface metamagnetic transitions in layered antiferromagnets, but also provides experimental support for realizing quantized transport in ultra-thin films of \ce{MnBi4Te7} and other natural superlattice topological magnets. 
\end{abstract}

\keywords{surface spin-flip transition, MFM, antiferromagnetic topological insulator}
\maketitle

Broken time reversal symmetry and topological band structure are the key ingredients for many interesting phenomena, such as the quantum anomalous Hall (QAH) effect and the topological magnetoelectric effect \cite{doi:10.1126/science.1234414,PhysRevLett.120.056801}. Although the QAH effect has been demonstrated in ferromagnetic topological insulator (TI) thin films, the inherent disorder from doping results in inhomogeneity that limits the quantization to sub-kelvin temperatures \cite{doi:10.1126/sciadv.1500740,Chang2015,PhysRevLett.114.187201}. Intrinsic magnetic TIs provide an alternative approach to combine magnetism and topological band structure in stoichiometric compounds. For example, the Z$_2$ topological index in A-type antiferromagnets is protected by the symmetry of alternating ferromagnetic layers \cite{PhysRevB.81.245209}.

\ce{MnBi2Te4} is the first tangible candidate for an antiferromagnetic-TI (AFM-TI) \cite{PhysRevLett.122.107202,Otrokov2019,doi:10.1126/sciadv.aaw5685}. The observation of quantum transport in exfoliated flakes provides strong evidence of QAH and axion insulator states in zero magnetic field \cite{doi:10.1126/science.aax8156,Liu2020}, though it remains controversal \cite{Ovchinnikov2021}. Indeed, high-resolution angle-resolved photoemission spectroscopy (ARPES) reports gapless Dirac surface states, suggesting that the surface spin configuration is different from the out-of-plane A-type AFM order in bulk \cite{PhysRevX.9.041040,PhysRevX.9.041038,PhysRevX.9.041039,PhysRevB.101.161109}. Previous magnetic force microscopy (MFM) studies by some of us, however, confirmed that the A-type antiferromagnetic order persists to the surface layer of \ce{MnBi2Te4}, in agreement with recent ARPES measurements \cite{PhysRevLett.125.037201,PhysRevLett.125.117205}. The robust A-type antiferromagnetic order is further corroborated by the observation of the long-sought surface spin-flop transition \cite{PhysRevLett.125.037201}.

In spite of mounting evidence of the robust A-type AFM order, it is possible that surface relaxation is limited to the very top layer and strictly follows morphology of surface steps so that it escapes the MFM observation. This scenario, however, requires an abrupt transition from ordered to relaxed magnetic states within the septuple layer beneath each step edge. This is physically unlikely given the strong intralayer exchange coupling \cite{PhysRevLett.124.167204}. If it is true, a further reduction of interlayer coupling by increasing the interlayer separation would favor a stronger surface relaxation effect, which can be visualized by magnetic imaging. The natural superlattice compounds \ce{MnBi2Te4-(Bi2Te3)}$_n$ provide perfect system to test such a hypothesis. In these systems, n layers of \ce{Bi2Te3} are inserted between \ce{MnBi2Te4} layers, dramatically reducing the interlayer coupling without much impact on the uniaxial anisotropy \cite{Hu2020,doi:10.1126/sciadv.aax9989,PhysRevMaterials.4.054202}. Thus, the metamagnetic transition becomes a spin-flip transition in \ce{MnBi4Te7} and \ce{MnBi6Te10} single crystals \cite{Hu2020,doi:10.1126/sciadv.aax9989,PhysRevMaterials.4.054202,Klimovskikh2020}. ARPES measurements observed gapless Dirac surface states on the \ce{MnBi2Te4} termination, again suggesting strong surface relaxation of the A-type AFM order \cite{Hu2020,PhysRevX.10.031013}. Therefore, it is imperative to probe the surface magnetism of the \ce{MnBi2Te4} termination in \ce{MnBi4Te7}. It is also interesting to find out whether there is a surface spin-flip transition proceeding the bulk one, which has been predicted theoretically but has evaded experimental observations \cite{R2004}.

In this letter, we report that the A-type AFM order persists to the surface \ce{MnBi2Te4} termination, as illustrated by the termination dependence of the magnetic signal observed by MFM, excluding the previous proposed surface relaxation of the A-type AFM order \cite{Hu2020,PhysRevX.10.031013}. In addition, we discover a first-order spin-flip transition on the \ce{MnBi2Te4} exposed surface that precedes the bulk spin-flip transition, in excellent agreement with a revised Mills’ model \cite{PhysRevLett.20.18,PhysRevLett.72.920,PhysRevLett.125.037201}. A quantitative analysis further suggests enhanced surface magnetization on the \ce{MnBi2Te4} termination, indicating robust two-dimensional ferromagnetism could exist in the single-layer limit \cite{PhysRevX.11.011003}. Therefore, \ce{MnBi4Te7} is promising material platform for achieving high temperature quantized transport in the thin film limit.

\begin{figure}[htbp]
    \centering
    \includegraphics[width=\columnwidth]{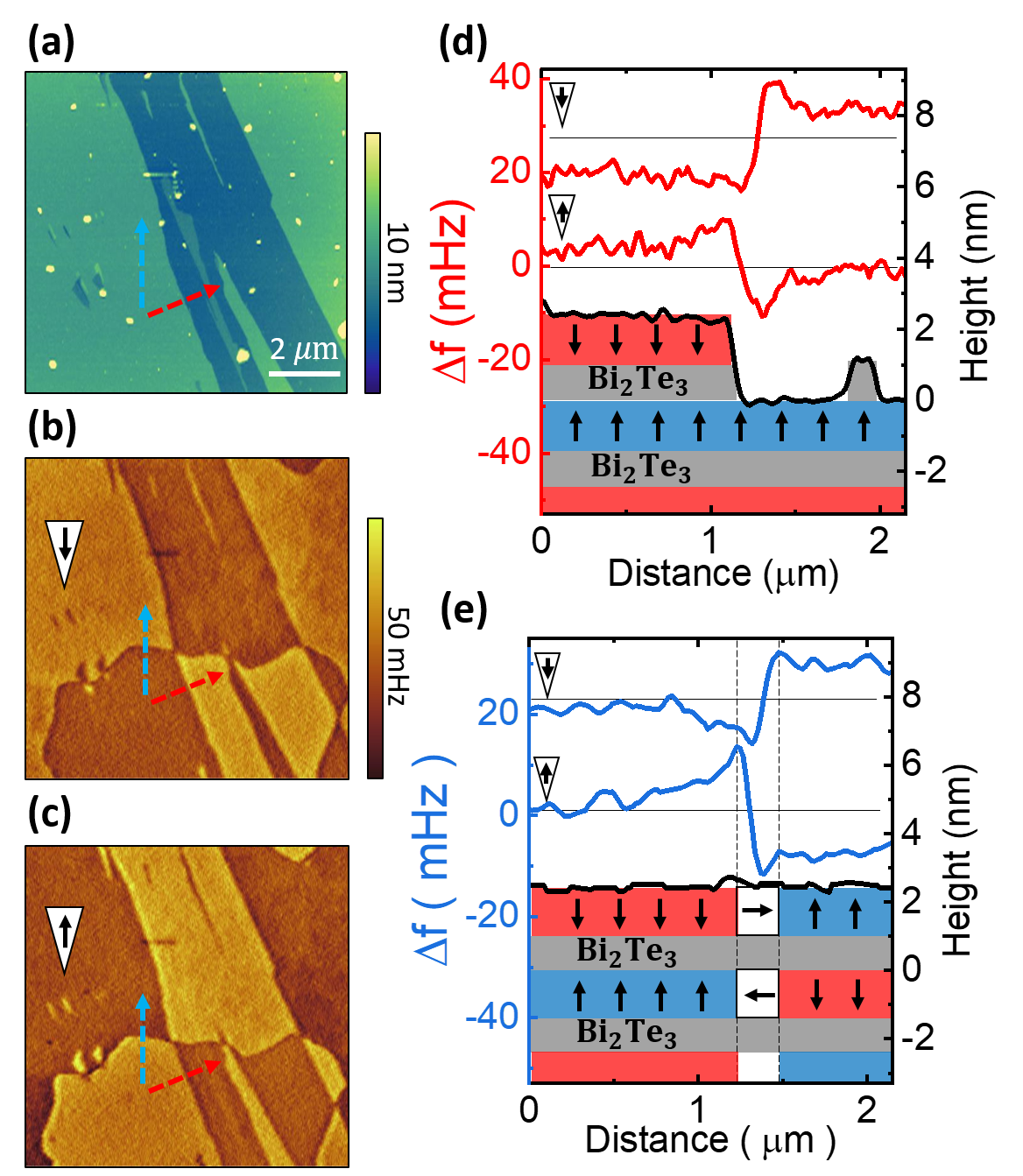}
    \caption{(Color Online) (a) Topographical image (5.5~K) measured on a cleaved surface of a \ce{MnBi4Te7} single crystal. (b),(c) MFM images (5.5~K, 0~T) measured at the same location as in a with negatively and positively polarized tip, respectively. The domain contrast is reversed as the tip moment is flipped, confirming that the magnetic contrast comes from the sample stray field. A curvilinear domain wall crossing the septuple layer (SL) and quintuple layer (QL) steps is visible in the MFM images. (d),(e) Line profiles of the topographical image (black) and MFM images (red and blue) along the red and blue arrow in (b) and (c). Schematics of crystal and magnetic stacking are sketched under the topographical line profiles. Red and blue layers represent the ferromagnetic SLs with moments pointing down and up, respectively. The gray layers are QLs. The frequency shift line profile is plotted across a SL plus QL step in (d) while it is plotted across the domain wall on a flat SL layer in (e).}
    \label{fig:1}
\end{figure}

\ce{MnBi4Te7} single crystals are grown out of \ce{Bi2Te3} flux \cite{supp}. Samples are cleaved in ambient condition to expose fresh surfaces before being mounted to a cryogenic MFM \cite{PhysRevLett.125.037201,Sass2020}. Because of the natural superlattice structure, the surface terminates at either \ce{MnBi2Te4} septuple layer (SL) or \ce{Bi2Te3} quintuple layer (QL). Figure~\ref{fig:1} shows the typical topography of a cleaved surface of \ce{MnBi4Te7} single crystal \cite{supp}. A trench with step height $\sim$2.4 nm cut through the field of view. The step height is approximately the c-axis lattice constant, indicating that it consists of a \ce{MnBi2Te4} SL and a \ce{Bi2Te3} QL. There are also a few islands inside the trench. Along the red arrow in Fig.~\ref{fig:1}(a), there is an island with height of $\sim$1.1~nm, indicating it is a QL (\ce{Bi2Te3}). Therefore, the majority of the surface is the \ce{MnBi2Te4} termination. Figure 1d shows the corresponding topographical line profile with a cartoon of the stacking order.

Figure~\ref{fig:1}(b) and (c) shows the MFM image measured with opposite tip moments at 5.5~K in zero external magnetic field. The tip moment is reversed by applying a 0.1~T external field, which is small enough without affecting the domain pattern but large enough to reverse the MFM tip moment. The bulk spin-flip transition is $\sim$0.13~T while the coercive field of MFM tip moment is $\sim$0.04~T \cite{supp}. The reversal of the MFM contrast with tip moment orientation confirms the magnetic signal is from the stray field of the sample. A curvilinear domain wall separating antiphase domains cuts across the trench. The MFM contrast reverses across the domain wall (blue arrow) or across the step on the same side of the domain wall (red arrow). The alternating MFM signal across both the domain wall and the step confirms that the out-of-plane A-type AFM order persists all the way to the surface \ce{MnBi2Te4} layer. The topographic and MFM line profiles with the corresponding magnetic structures are shown in Fig.~\ref{fig:1}(d) and (e). The absence of magnetic contrast on \ce{Bi2Te3} island suggests its magnetic signal is negligible at 5.5~K and higher temperatures even though Mn$_\textrm{Bi}$ defects in \ce{Bi2Te3} carry magnetic moments \cite{PhysRevB.81.195203}. Thus, the \ce{Bi2Te3} layer behaves as a non-magnetic spacer in \ce{MnBi4Te7}. Thus, the magnetic contrast observed in this work originate from the magnetic order in the \ce{MnBi2Te4} layers. At lower temperatures, the magnetism in \ce{Bi2Te3} layers could interact with the AFM order in \ce{MnBi2Te4} layers, which might be related to the substantial hysteresis loop of the bulk spin-flip transition \cite{Hu2020,doi:10.1126/sciadv.aax9989,PhysRevMaterials.4.054202,Hu_2021}. The persistence of out-of-plane A-type AFM order suggests that \ce{MnBi4Te7} is a perfect system to explore the surface metamagnetic transition, similar to the surface spin-flop transition observed in \ce{MnBi2Te4} single crystals \cite{PhysRevLett.125.037201}. 

\begin{figure}[htbp]
    \centering
    \includegraphics[width=\columnwidth]{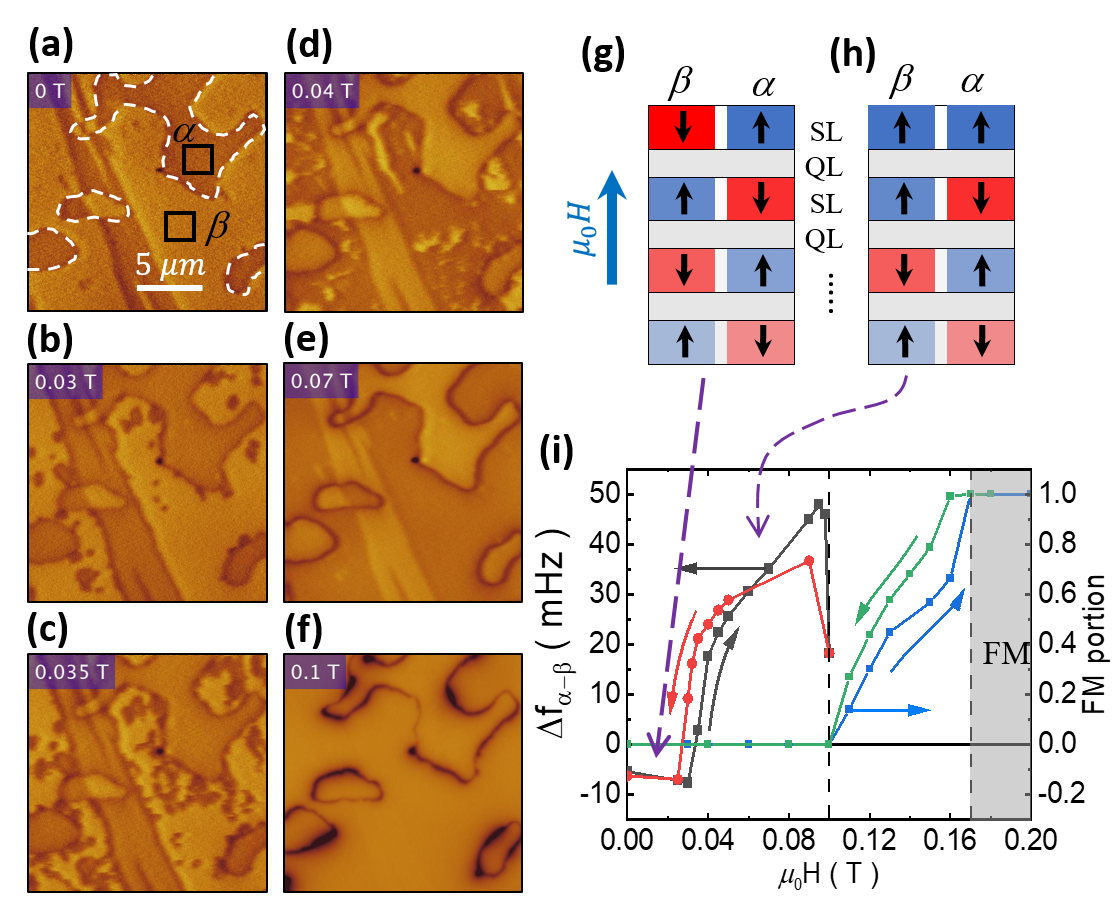}
    \caption{(Color Online) (a-f) Selected MFM images taken at 5.5~K with increasing magnetic fields, which are labeled on the corner of each image. The color scales from (a) to (f) are 0.06, 0.11, 0.12, 0.12, 0.26 and 3~Hz, respectively. The AFM domain walls are traced out with dashed lines in (a). The boxed areas $\alpha$ and $\beta$ in a corresponds to parallel and anti-parallel surface, respectively. 
     (g)/(h) Magnetic structures of areas $\alpha$ and $\beta$ before/after the surface spin-flip transition, respectively. (i) $H$-dependence of MFM contrast between areas $\alpha$ and $\beta$, and that of forced ferromagnetic (FM) domain population during the bulk spin-flip (BSF) transition (between 0.1 and 0.17~T). The colored arrows indicate the field sweeping direction. Above 0.17~T, the system is in the FM state.}
    \label{fig:2}
\end{figure}

In \ce{MnBi4Te7}, the insertion of \ce{Bi2Te3} layer dramatically reduces the interlayer exchange interaction without affecting the uniaxial magnetic anisotropy. Therefore, the metamagnetic transition becomes a spin-flip transition \cite{Hu2020,doi:10.1126/sciadv.aax9989,PhysRevMaterials.4.054202}. The bulk spin-flip (BSF) transition field is $\mu_0H_\textrm{BSF} \approx$ 0.13~T at 5.5~K, in good agreement with recent MFM studies \cite{Hu_2021, supp}. 
Thus, the surface \ce{MnBi2Te4} layer with antiparallel moment is expected to undergo a surface spin-flip transition before the bulk transition because of reduction of Weiss field due to the missing of half nearest neighbors \cite{R2004}. Figure~\ref{fig:2}(a-f) show selected MFM images measured at various out-of-plane magnetic field after 0.01~T field cooling through the Néel temperature ($T_\textrm{N}\approx$ 13~K) \cite{PhysRevMaterials.4.054202}. Positive field value indicates the direction of field is up. Curvilinear domain walls separating $\alpha~ (\uparrow\downarrow\uparrow\downarrow)$ and $\beta~ (\downarrow\uparrow\downarrow\uparrow)$  antiphase domains are highlighted in Fig.~\ref{fig:2}(a). The spin configurations of two types of antiphase domains are illustrated in Fig.~\ref{fig:2}(g). As magnetic field is increased to 0.03~T, a few bubble-like features with dark contrast appear only on antiparallel surfaces ($\beta$ domains above the trench) as shown in Fig.~\ref{fig:2}(b), indicating a metamagnetic transition that proceeds the bulk spin-flip transition. More dark features nucleate and expand with further increasing magnetic field. The dark contrast take over the whole antiparallel surface at 0.07~T as shown in Fig.~\ref{fig:2}(b-e). After that, the magnetic contrast between two surface terminations (parallel and antiparallel) is reversed, as summarized in Fig.~\ref{fig:2}(i) \cite{supp}. 
Since the transition only happens on the antiparallel surface, it is the long-sought surface spin-flip (SSF) transition \cite{R2004}. The first-order nature of the SSF transition is further corroborated by the small hysteresis between increasing and reducing field results shown in Fig.~\ref{fig:2}(i). Note that the magnetic contrast of AFM domains is much ($\sim$1000 times) weaker than the contrast between the AFM phase and the forced ferromagnetic phase in the BSF transition, further corroborating observed domain process is the transition of the surface layers \cite{supp}. 
The magnetic structure of the boxed region before and after the SSF transition is shown in Fig.~\ref{fig:2}(g) and~\ref{fig:2}(h). Consistently, the same SSF transition is observed on the opposite termination ($\alpha$ domains above the trench) with negative (downward) magnetic fields \cite{supp}, confirming the SSF transition is an intrinsic phenomenon of the surface layer with moment antiparallel to the external field. Similar to prior MFM studies of \ce{MnBi2Te4}, the magnetic contrast of domain walls linearly increases with increasing magnetic field at small fields, suggesting a susceptibility contrast mechanism \cite{PhysRevLett.125.037201,https://doi.org/10.48550/arxiv.2112.02320}.

The observed surface spin-flip transition field ($\mu_0H_\textrm{BSF}$) is approximately 1/4 of that of the bulk one. We observed similar ratio in different samples with slightly different transition temperature and fields \cite{supp}. 
Furthermore, the ratio doesn’t vary much for $T <$ 10~K ($\sim$80\% of $T_\textrm{N}$), suggesting that the SSF transition follows the BSF one. To understand the mechanism of SSF transition, we performed analysis using a revised Mills’ model in the high anisotropy limit ($K/J\gg$ 1) . Here $K$ is the uniaxial anisotropy energy, and $J$ is the exchange energy \cite{PhysRevLett.98.106803,PhysRevB.50.3931,PhysRevB.81.245209}. In comparison, the previous modeling of the surface spin-flop transition in \ce{MnBi2Te4} is in the low anisotropy limit ($K/J\ll$ 1) \cite{PhysRevLett.125.037201}. Therefore, in contrast to the claim of recent MFM studies \cite{https://doi.org/10.48550/arxiv.2112.02320}, there is no surface spin-flop transition in \ce{MnBi4Te7}. 

\begin{figure}[htbp]
    \centering
    \includegraphics[width=\columnwidth]{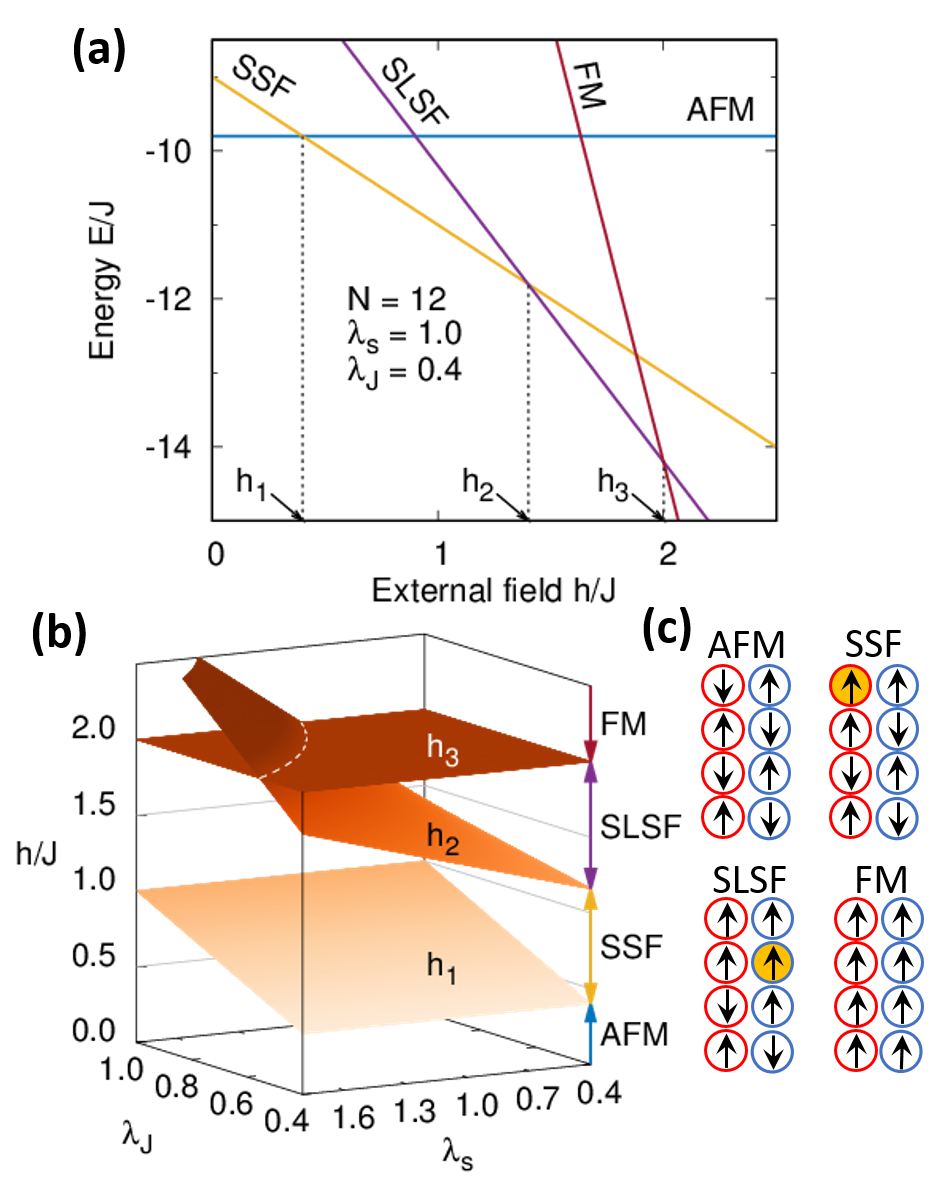}
    \caption{(Color Online) (a) Total energy as a function of exchange field, for the four magnetic phases with twelve spin-lattice sites ($N = 12$) and reduced surface exchange coupling ($\lambda_J=0.5$). First order phase transitions are expected at the crossing points $h_1$, $h_2$ and $h_3$, which are surface spin-flip (SSF), second-layer spin-flip (SLSF), and bulk spin-flip transitions, respectively. (b) The crossing points with respect to the surface exchange coupling $\lambda_J$ and the surface spin moment $\lambda_S$. The SLSF phase can appear only when the surface parameters are reduced as $\lambda_J\cdot\lambda_S<1$. (c) Schematics illustration of the surface spin-flip states where red and blue represent antiphase domains. Assuming the external field points up, SSF (SLSF) occurs at the first (second) layer of antiparallel (parallel) surface as highlighted by orange filling.}
    \label{fig:3}
\end{figure}

The Mills’ model is effectively a one-dimensional spin chain model where each spin represents the magnetic moment of each layer in A-type AFMs \cite{PhysRevLett.20.18}. The strong uniaxial anisotropy ($K/J\gg$ 1) forces all spins to align on the vertical easy axis. In this limit., the anisotropy term can
be omitted from the original model, the total energy is simplified to 
$\displaystyle     E=J\sum_{i=1}^{N-1} \boldsymbol{S}_i \cdot\boldsymbol{S}_{i+1}-\sum_{i=1}^{N}\boldsymbol{S}_i\cdot\boldsymbol{h}$. 
Thus, the total energy of AFM ground state is, $E_\textrm{AFM}=-(N-3+2\lambda_J\lambda_S)J$, where $\lambda_J$ is the ratio of the revised surface exchange coupling to the bulk one, and $\lambda_S$ is the ratio of the revised surface spin moment to that in bulk \cite{PhysRevB.81.245209}. Here, $\lambda_J$ and $\lambda_S$ are phenomenological parameters that characterize the effect of surface relaxation. The $E_\textrm{AFM}$ is independent of external field because of compensated magnetic moments. If the surface layer or the second layer moment reverses, the Zeeman energy gain of the uncompensated moments would result in first-order transitions. Figure~\ref{fig:3}(a) shows the total energies of four spin states for $\lambda_J\lambda_S<$ 1. The schematics are shown in Fig.~\ref{fig:3}(c).  First-order phase transitions occur at threshold fields of $h_1=\lambda_JJ$, $h_2=(1+\lambda_J\lambda_S)J$, and $h_3=2J$, where the lowest total energy evolves from AFM state to the force ferromagnetic state via the SSF and second-layer spin-flip (SLSF) states. Note that the $h_1$ only depends on $\lambda_J$. As discussed earlier, the ratio $H_\textrm{SSF}/H_\textrm{BSF}$ is approximately 1/4, which is $h_1/h_3$ in our model. Thus, the ratio of revised surface exchange can be estimated as $\lambda_J=2h_1/h_3\approx0.5$. In other words, the exchange coupling between the surface layer and the next layer is approximately half of the value in bulk of \ce{MnBi4Te7}, probably due to surface relaxation effect. Interestingly, the revised Mills’ model also predicts a second-layer spin-flip transition ($h_2$) between the SSF transition ($h_1$) and the bulk transition ($h_3$) for $\lambda_J\lambda_S< 1$, Otherwise the total energy of the SLSF phase is always above the lowest energy states \cite{supp} so that the system undergoes a phase transition from SSF to FM states above a threshold field of $h_\textrm{SSF-FM}=2[(N-3+\lambda_J\lambda_S)/(N-2)]J$, which approaches $h_3$ in the bulk limit ($N\rightarrow\infty$). Experimentally, no signature of the SLSF transition is observed before the bulk spin-flip transition begins ($\sim$0.1~T), indicating $\lambda_J\lambda_S\geq1$. However, the SLSF transition might be hidden by the relatively broad ($\sim$0.07~T) BSF transition.

\begin{figure}[htbp]
    \centering
    \includegraphics[width=0.98\columnwidth]{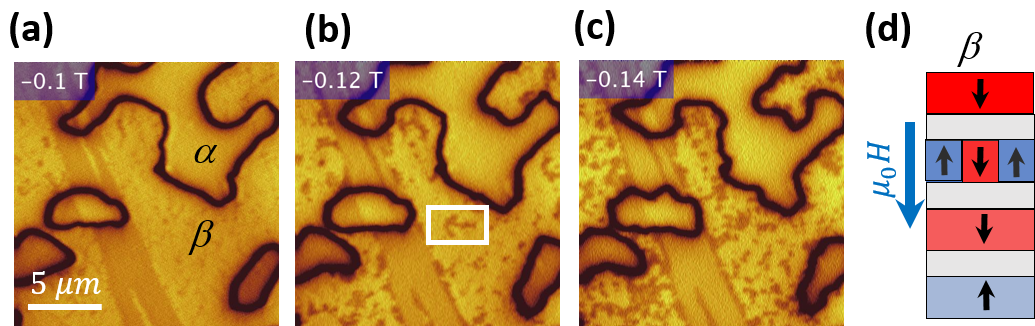}
    \caption{(Color Online) (a)-(c) MFM images taken at $-0.09$~T after external magnetic field was increased to the set values labeled at upper-left corners. The color scale is 0.2~Hz. The field “annealing” results illustrate partial second layer spin-flip (SLSF) transition on the parallel surface.  (d) Schematic illustration of SLSF phase. The magnetic moment in the second SL of $\beta$ domain is partially flipped, which corresponds to the dark patches as highlighted in the white box in (b).}
    \label{fig:4}
\end{figure}

To explore whether the SLSF transition is overshadowed by the BSF transition, we performed field “annealing” experiments by carefully increasing magnetic field to induce partial but reversible BSF transition ($\mu_0H\leq$ 0.14~T) \cite{supp}. 
For negative field, the surface of $\alpha$ domains undergo SSF transition at $-0.033$~T \cite{supp}. Figure~\ref{fig:4}(a) shows the MFM image taken at $-0.09$~T after sweeping magnetic field to -0.1~T. Interestingly, numerous small patches with dark contrast appear on the parallel surface indicating a partial SLSF transition. As shown in Fig.~\ref{fig:4}(b) and (c), more fraction of $\beta$ domains undergo partial SLSF transition with increasing fraction of dark patches after $-0.12$ and $-0.14$~T field “annealing”. So the SLSF transition field is very close to that of BSF transition, {\it i.e.}, $h_2\approx h_3$, indicating $\lambda_S\approx2$. Therefore, these results suggest that the moment of the surface \ce{MnBi2Te4} layer is larger than the bulk value. 
The enhanced surface moment indicates that a robust 2D ferromagnetism could persist in the single layer \ce{MnBi2Te4} limit, which is favorable for exploring the quantum transport in thin films or flakes of \ce{MnBi4Te7} and related superlattice compounds. 

In summary, we discover the SSF transition in AFM-TI \ce{MnBi4Te7}, in good agreement with a revised Mills’ model. Furthermore, we observed a partial SLSF transition, suggesting enhanced magnetic moment in the surface \ce{MnBi2Te4} layer. The alternating domain contrast across the domain wall or step edge observed in \ce{MnBi4Te7} unambiguously confirms the persistence of A-type AFM order to the surface \ce{MnBi2Te4} layer. The discovery and direct visualizing of SSF transition paves the way for exploring surface or 2-dimensional magnetic states of functional AFMs for spintronic applications \cite{RevModPhys.90.015005}. Moreover, the robust ferromagnetism in the single-layer limit opens door to realize QAH or axion insulator states in the ultra-thin films of the natural superlattice \ce{MnBi4Te7} and related compounds \cite{Hu2020,doi:10.1126/sciadv.aax9989,PhysRevLett.123.096401}.

\begin{acknowledgments}
The MFM studies at Rutgers is supported by the Office of Basic Energy Sciences, Division of Materials Sciences and Engineering, US Department of Energy under Award numbers DE-SC0018153. The simulation efforts is supported by NSF grant DMR-1954856. Work at ORNL was supported by the US Department of Energy, Office of Science, Basic Energy Sciences, Materials Sciences and Engineering Division.
\end{acknowledgments}

\bibliographystyle{apsrev4-2}
\bibliography{SSF}

\end{document}